# THREE NEARBY K-GIANTS WITH PLANETS: ACCURATE DETERMINATION OF BASIC PARAMETERS, INCLUDING AN ANALYSIS OF METALLICITY BASED ON Fe I LINES

L. S. Lyubimkov, D. V. Petrov, and D. B. Poklad

*Fundamental parameters, including the effective temperature $T_{eff}$, surface gravity log$g$, mass M, luminosity L, radius R, and age t, are determined for three bright, nearby K-giants, β Gem (K0 III), μ Leo (K2 III), and α Tau (K5 III). It is notable that around all three stars the giant planets have been found. Our results are compared with published high-precision data for benchmark stars included in the* Gaia *project. Very good agreement was obtained for all three giants with these data for the basic parameters $T_{eff}$, log$g$, and M, although our technique of their determination was simpler. Special attention was devoted to analy-sis of the Fe I lines which are the basis of a simultaneous determination of the metallicity index [Fe/H] and microturbulent parameter $V_t$. The equivalent widths W of the Fe I lines are automatically measured from published spectra for the benchmark stars. An analysis of Fe I lines from the list of "golden lines" selected in a study of benchmark stars led to the conclusion that the excitation potential $E_l$ of the low level of the lines plays an substantial role in determining [Fe/H] and $V_t$. It is shown that in the case of the early K-giants β Gem and μ Leo with temperatures $T_{eff}$ between 4400 and 4900 K for lines in the range of W from 100 to 300 mÅ there is a dependence of the [Fe/H] and $V_t$ values on $E_l$. This dependence was not taken into account earlier is studies of the K-giants, in particular in the Fe I lines analysis of the benchmark stars. It is shown that a correct accounting for it leads an ambiguity in the determination of [Fe/H] and $V_t$ for β Gem and μ Leo. In the case of the coolest K-giant α Tau (Aldebaran) with a temperature $T_{eff}$ = 3920K, this effect seems to be less pronounced. Recommendations are given for the*

---

Crimean Astrophysical Observatory, RAN, Russia; e-mail: lyub@craocrimea.ru

*Fe I lines selection for [Fe/H] and $V_t$ determination. It is sonfirmed for the example of the three stars studied here that the non-LTE effects in the Fe I lines for the K-giants with normal metallicity are very minor, so they cannot be the reason of the revealed ambiguity in [Fe/H] and $V_t$ values. The low ratios of the carbon $^{12}C/^{13}C$ and oxygen $^{16}O/^{17}O$ isotopes confirm that all three giants passed the phase of deep convective mixing.*

## 1. Introduction

The cold giants are visible from large distances, which, together with their large numbers, makes these stars attractive objects for many studies of the properties of the galaxy. These studies include, in particular, analyses of the radial gradient of the abundances of chemical elements in the galaxy. In one example [1], a study of 304 red giants in 29 open clusters yielded the gradients in the amounts of a number of elements in the galactic disk at distances of 8-15 kpc from its center.

For the studies done at the Crimean Astrophysical Observatory (CrAO), these kinds of stars are of interest from several standpoints. First, for the ongoing studies at CrAO of the properties of cold giants and supergiants that are rich and super-rich in lithium (this rare type of star cannot be explained in terms of the standard theory of stellar evolution). Second, for the studies begun at CrAO of K-giants near which planets have been detected (this paper is an example). Third, for a planned study of cold giants which are treated as possible successors to magnetic CP stars. In all these cases, precise data on the fundamental parameters of these stars and their chemical composition are needed; only with such a reliable foundation is it possible to determine any distinctive properties of these stars and establish their nature.

The first of these two problems concerning giants and supergiants that are rich and super-rich in lithium has been discussed in some detail in a review [2]; some details on light elements are also given in Ref. 3. As noted there [2,3], most lithium-rich stars and all stars that are super-rich in lithium cannot be explained in terms of the standard theory of stellar evolution. These stars are very few in number (1-3% of the total number of FGK-giants) and they all have masses $M < 6 M_\odot$. The hypothesis that these stars capture giant planets with the mass of Jupiter or greater has been under very active discussion in recent years. This capture may lead to a significant increase in the lithium content in a star's atmosphere. Phenomena of this kind cannot be regarded as a great rarity; according to a current estimate [4], the rate at which planets are incident on red giants is roughly 3 events per year per galaxy similar to our own.

While in the first problem regarding cold lithium-rich cold giants we can only speak of the hypothetical existence of planets near these stars, the second problem involves red giants with planets that have already been discovered around them. A study of the parameters of about ten bright, nearby K-giants near which planets with the mass of Jupiter or more have been discovered in recent years has been started at CrAO. This paper is part of such a study.

The third problem is related to searches for possible successors to magnetic chemically peculiar stars (CP-stars)

in classes A, F, and G which are in the main sequence (MS) stage, among objects in the next stage stage of evolution – the FGK-giant stage. The search for successors of CP-stars among cold giants must be based on the relatively small magnetic fields (up to 100 G) observed in a few stars of this type and possible (residual) excesses of heavy elements.

In order to solve these kinds of problems it is necessary, first of all, to determine the fundamental (basic) parameters of these stars, including their effective temperature $T_{eff}$, surface gravitation $g$ (usually given on a logarithmic scale, log$g$), and metallicity index [$Fe/H$], which is found in terms of the iron abundance $\log\varepsilon(Fe)$, with sufficient accuracy. The latter quantity is given on a standard scale, with $\log\varepsilon(H) = 12.00$ assumed for hydrogen.

We recall that the metallicity index is defined by [Fe/H] $= \log\varepsilon(Fe) - \log\varepsilon_\odot(Fe)$, i.e., it is given by the difference in the abundances of iron in a planet and the sun; we take $\log\varepsilon_\odot(Fe) = 7.50$ [5] for the solar abundance. The microturbulence parameter $V_t$, upon which the analysis of the abundances of all the other chemical elements depends, is found at the same time as [$Fe/H$]. We have evaluated the accuracy of the method used at CrAO for determining the basic parameters of the K-giants by comparing our results with Refs. 6-8.

All of the above parameters were studied in detail in Refs. 6-8 for 34 "benchmark stars" in classes F, G, and K of interest for calibration of the results of the *Gaia* project in which observations (mainly astrometric) were made of about a billion stars in our galaxy. The metallicity index [Fe/H] for these 34 stars was studied in detail in Ref. 6, the parameters $T_{eff}$ and log$g$ in Ref. 7, and the abundances of a number of elements from Mg to Ni in Ref. 8. Three stars in the "benchmark stars" list, specifically the giants β Gem (K0 III), μ Leo (K2 III), and α Tau (K5 III), are in the list of nearby K-giants with planets planned for studies at CrAO.

In this paper we determine the basic parameters $T_{eff}$, log$g$, [$Fe/H$], and $V_t$, as well as the mass $M$, luminosity $L$, radius $R$, and age $t$ for these giants β Gem, μ Leo, and α Tau. We evaluate the accuracy of our technique by comparison with highly accurate values of the basic parameters obtained in Ref. 6 and 7. Since we shall often refer to these data, for brevity we denote data from Refs. 6 and 7 as BSD (Benchmark Star Data).

For us it is important that the spectra of the three K-giants used in the BSD are accessible. Thus, we can base our work on the same observational spectral data as in the BSD, in particular when analyzing the lines of Fe I. The latter is the subject of special attention, since it led to some extraordinary conclusions in determining the parameters [$Fe/H$] and $V_t$.

**2. Some data on the K-giants studied here**

Table 1 lists some data on the three stars, including their HR and HD numbers, visible magnitude, and spectral subclass. The observed rotational velocities $V\sin i$ shown here are taken from Ref. 9, where giants in the distance range $d < 100$ pc of interest to us were examined. The parallaxes π and corresponding distances $d = 1/\pi$ are based on data from Hipparcos [10]. Table 1 shows that the distances of these giants from the sun range from 10 to 38 pc; that is, bright and very close stars are being examined.

All three of these stars are distinctive in that planets have been discovered for all of them in recent years as the result of many years of observations. The masses of these planets, more precisely the values of $m\sin i$, which represent a lower limit for the mass $m$, are 2.6, 2.4, and 6.5 times the mass of Jupiter, respectively, for β Gem [11], μ Leo [12], and α Tau [13]. The rotation periods of the planets around these stars are quite large at 596.6, 357.8, and 629.0 days, respectively.

TABLE 1. Some Data for the Three K-Giants Studied Here

| Star | HR | HD | $V$, mag | Sp | $V\sin i$, km/s | $\pi$, mas | $d$, pc |
|---|---|---|---|---|---|---|---|
| β Gem | 2990 | 62509 | 1.14 | K0 III | 2.8 | 96.54±0.27 | 10 |
| μ Leo | 3905 | 85503 | 3.88 | K2 III | 4.5 | 26.28±0.16 | 38 |
| α Tau | 1457 | 29139 | 0.86 | K5 III | 4.3 | 48.94±0.77 | 20 |

## 3. Fundamental parameters

The effective temperature $T_{eff}$ and acceleration of gravity $\log g$ are fundamental (or basic) parameters of the stars. They are known to be closely related to two other fundamental quantities, the mass $M$ and luminosity $L$. A determination of $T_{eff}$ and $\log g$ for a star precedes the analysis of its chemical composition and the accuracy with which elemental abundances are determined depends on the accuracy of these parameters.

**3.1. Determination of the effective temperature $T_{eff}$.** It is well known that the spectra of cold stars depend on the effective temperature $T_{eff}$. Thus, this parameter plays an important role in studies of K-giants. In particular, the Fe I lines are particularly sensitive to $T_{eff}$ in an analysis of the abundance of iron in cold stars, which is the focus of this paper(they depend on $\log g$ relatively weakly).

There is a direct method for estimating the effective temperature $T_{eff}$ that follows immediately from the definition of this quantity: $H_r = \sigma_r T_{eff}^4$, where $H_r$ is the total (integrated) radiative flux from the star and $\sigma_r$ is the Stefan constant. When the direct method is used it is necessary to know the distance $d$ to the star, the observed bolometric flux, and the angular diameter $\theta$. Knowing the distance $d$, which is fairly precisely determined for nearby stars from the measured parallax, one can proceed from the angular diameter $\theta$ to the star's linear radius R. Note that $\theta$ can be measured only for fairly nearby stars, usually with interferometers (less often by the method of star occultation by the moon). The uncertainty in measuring $\theta$ is the main source of errors in determining $T_{eff}$ by the direct method.

For two of the giants we have studied, α Tau and β Gem, the direct method of measuring $T_{eff}$ has been used in Ref. 7. For μ Leo, the most distant of the three stars and the one with the smallest angular diameter (see Table

TABLE 2. Determination of the Effective Temperature $T_{eff}$, K

| Star | [Fe/H] [16] | Q | $T_{eff}$ LP'14 | $T_{eff}$ [16] | $T_{eff}$ [17] | $T_{eff}$ (assumed) |
|---|---|---|---|---|---|---|
| β Gem | 0.13 | 0.140 | 4810 | 4821 | 4850 | 4830 |
| μ Leo | 0.42 | 0.494 | - | 4471 | 4480 | 4475 |
| α Tau | -0.01 | 0.748 | 3950 | 3903 | 3910 | 3920 |

4 of Ref. 7), θ was estimated indirectly in Ref. 7 using Arcturus as a comparison star. We compare these exact values of $T_{eff}$ for the three stars with the values of $T_{eff}$ found by our method in the following.

When stars are sufficiently distant, the direct method of determining $T_{eff}$ becomes inapplicable. In this case, for cold stars in the region $T_{eff}$ = 3900 - 5000 K we are considering the most reliable methods of determining $T_{eff}$ are photometric. One of these reliable and popular methods is the infrared flux method (IRFM). As a first step we used the photometric method developed by Lyubimkov and Poklad [14], referred to below as LP'14, to determine $T_{eff}$ for G- and K-giants and supergiants, and calibrated on the basis of IRFM data. This method is based on using the indices $Q$ and $[c_1]$ in the UBV and $uvby$ photometric systems, respectively. It is important that both these indices be free of the influence of interstellar absorption. For our stars, only the index $Q$ can be used, since the Q-method operates within a fairly narrow range of $T_{eff}$ = 3800 - 5100 K, which again corresponds to the K-giants being studied here. Note that the index $[c_1]$ can be used for somewhat hotter stars with temperatures $T_{eff}$ = 4900 - 5500 K.

As an example of the rather high accuracy of the Q-method, the authors of LP'14 presented a determination of $T_{eff}$ for Arcturus ( α Boo), a bright and very close K-giant, which has been the object of many studies. Here it was noted that this star has a reduced metallicity [Fe/H] = – 0.5. The Q-index was used to find a temperature of $T_{eff}$ = 4262 ± 20 K, which differed by only 24 K from the value $T_{eff}$ = 4286 ± 30 K obtained in Ref. 15 from the energy distribution in the spectrum of Arcturus over a wide range from 0.44 to 10 mm.

The LP'14 method yields the relationship between the effective temperature $T_{eff}$ and the observed Q-index for metallicities of [Fe/H] = 0.0 and –0.5. Thus, when this method is used it is necessary to know (even if approximately) the parameter [Fe/H]. Table 2 lists the values of [Fe/H] from the paper by Luck [16]. We note that these values of [Fe/H] will be significantly refined in the following, especially for α Tau.

The index Q in the UBV photometric system is given by the formula $Q = (U-B) - 0.72(B-V)$. The observed UBV values for our bright giants are known to high accuracy; e.g., in the SIMBAD data base (http://simbad.u-strasbg.fr/simbad/sim-fid). Based on the values of [Fe/H] and Q in Table 2, we determined the effective temperature $T_{eff}$ for the giants β Gem and α Tau by the LP'14 method. For m Leo the metallicity [Fe/H] = +0.4 turned out to be too high, so the LP'14 method could not be used in that case.

As for determining $T_{eff}$ for the giant α Tau, it should be noted that the value [Fe/H] = – 0.01 given in Table

2 appears to be too high; the actual metallicity index for this star may be –0.3 or –0.4 dex (see below). In the range of $Q \sim 0.7$ where this star falls, the LP'14 method cannot be used when $[Fe/H] = -0.5$; however, for these values of Q it may be expected that the method should not yield significant differences between the cases of $[Fe/H] = 0.0$ and –0.5 (see Fig. 2 in Ref. 14). Thus, when estimating $T_{eff}$ for α Tau we used Eq. (1) obtained in LP'14 for $[Fe/H]=0.0$.

For our bright giants, $T_{eff}$ has been estimated a number of times by various authors. About 20-30 estimates of $T_{eff}$ are given for each of them in the SIMBAD database. We note again that for cold stars in the $T_{eff} = 3900-5000$ K region, the most reliable determinations of $T_{eff}$ are given by photometric methods in the $T_{eff}$ range if the direct method of estimating $T_{eff}$ cannot be used. In Table 2 we give the values of $T_{eff}$ found by Luck [16] using the same IRFM method upon which the LP'14 method is based. We have also used the estimates of $T_{eff}$ by McWilliam [17] for 671 giants in classes G and K based on 10 color indices. This paper is still being cited widely, so we give the values of $T_{eff}$ from this paper for our three stars.

Table 2 shows that three different sources of values for $T_{eff}$ based on different photometric methods yield similar results; the differences in $T_{eff}$ do not exceed 40 K. The last column of Table 2 gives the assumed value of $T_{eff}$, which is close to the average.

The resulting values of $T_{eff}$ vary from 4830 K for the hottest K0-giant β Gem (Pollux) to 3920 K for the coolest K5-giant α Tau (Aldebaran). In fact, this range overlaps the entire interval of typical values of $T_{eff}$ for K-giants (recall that the K-giants as a whole are subdivided into spectral subtypes from K0 to K5).

**3.2. Determining the acceleration of gravity log*g* and mass *M*.** We found the second fundamental parameter, the acceleration log*g* of gravity in the star's atmosphere, from the trigonometric parallax π by a method described by Lyubimkov, et al. [18,19]. Thanks to the high accuracy of the values of π obtained with the HIPPARCOS satellite [10] for relatively nearby stars, this method of determining log*g* can today be regarded as one of the most accurate for these stars. We note that the parallaxes were also used for determining log*g* in the BSD papers (see below).

The method described in Refs. 18 and 19 can be used for determining the mass *M* simultaneously with log*g*. For this it is necessary employ calculations of the evolutionary tracks of the stars for a number of values of *M*. In modern tracks, the initial content of metals *Z* is varied along with *M*. Thus, when determining log*g* and *M* for a particular star, the variations in the metallicity parameter [*Fe/H*] from the normal (solar) value can be taken into account.

In this work we initially used the evolutionary tracks of Claret [20] calculated for a normal metallicity Z=0.02. We have used these tracks in previous studies of supergiants and giants in classes A, F, G, and K [19]. Now we have used them to determine log*g* and *M* for the stars b Gem and m Leo.

For determining log*g* and *M* for α Tau it was necessary to account for the reduced (roughly by a factor of two) metallicity of this star. Thus, in this case we have used tracks from the same author [21] calculated for $Z = 0.01$. It should be noted that a determination of log*g* and *M* for such a cold giant as α Tau was extremely sensitive to the metallicity index [*Fe/H*] or *Z*, as well as to the effective temperature $T_{eff}$. Thus the errors in determining these parameters, as well as other values associated with the use of evolutionary tracks, turned out to be substantially higher for α Tau than for the hotter giants β Gem and μ Leo with normal metallicities (see below).

It should be noted that in the BSD the acceleration of gravity was found using the formula defining this quantity, i.e., $g = GM/R^2$, where M is the star's mass, R is its radius, and G is the gravitational constant. The linear radius R was found in terms of the measured angular diameter θ (see section 3.1) and the parallax π. Thus, in the BSD log$g$ was also found using the parallaxes, but in a different context than in our work.

**3.3. The basis parameters for the three K-giants and comparison with the results for the BSD.** The parameters $T_{eff}$, log$g$, and $M$ that we found are listed in Table 3. The BSD values [7] of these three parameters for the same three stars are also listed there.

Once $T_{eff}$, log$g$, and $M$ are known, the known formulas can be used easily to obtain the star's luminosity L and radius R. The values of L and R relative to the solar values are also listed in Table 3. In addition, here we give our estimates of the age $t$ derived from the evolutionary calculations of Claret [20, 21] mentioned above.

Table 3 shows that our values of the effective temperature. $T_{eff}$ for the three K-giants in the program are in well agreement with the BSD data. It is important that our analysis of $T_{eff}$ based on a photometric technique gives essentially the same results as the BSD obtained by the direct method, i.e., using measurements of the angular diameters of these nearby stars.

With regard to our values for log$g$, we note that the difference from the BSD values is less than 0.10 dex; i.e., the difference is within the limits of error for determining these values.

TABLE 3. Basis Parameters Found Here for the Three K-Giants Compared with BSD Data [7]

| Parameter | Source | β Gem | μ Leo | α Tau |
|---|---|---|---|---|
| $T_{eff}$, K | our work | 4830±30 | 4475±30 | 3920±40 |
| | BSD | 4858±60 | 4474±60 | 3927±40 |
| log$g$ | our work | 2.85±0.10 | 2.43±0.10 | 1.20±0.15 |
| | BSD | 2.90±0.08 | 2.51±0.11 | 1.11±0.19 |
| $M/M_\odot$ | our work | 2.3±0.2 | 1.6±0.2 | 1.2±0.4 |
| | BSD | 2.3±0.4 | 1.7±0.4 | 1.0±0.4 |
| log $L/L_\odot$ | our work | 1.65±0.04 | 1.78±0.06 | 2.66±0.14 |
| | BSD | 1.60±0.02 | 1.71±0.02 | 2.64±0.02 |
| $R/R_\odot$ | our work | 9.5±0.3 | 12.9±0.7 | 46.2±6.8 |
| Age, $10^9$ years | our work | 0.8±0.2 | 2.4±0.8 | 5.5±2.0 |

As for the masses M given in Table 3, it should be noted that evolutionary tracks from different authors were used for determining M in our paper and in the BSD. We used the calculations of Claret [20, 21], while the BSD employed two different types of evolutionary models: Padova [22,23] and Yonsei-Yale [24, 25]. The agreement between these two last sources was good: the difference in the values of M for the K-giants does not exceed 0.15 $M_\odot$ (see Fig. 3 of Ref. 7). It is important that, despite the use of different evolutionary tracks, our values of M and the data from BSD were in very good agreement (Table 3).

It can be seen from Table 3 that fairly good agreement was also obtained for the luminosities: the difference from the BSD in $\log L/L_\odot$ is comparable to the errors in determining this quantity.

The radii R and age t of the test stars are not given in the BSD. Our estimates, obtained using the calculations of Refs. 20 and 21, showed that $R/R_\odot$ equals 9, 13, and 46, while the age t equals 0.8, 2.4, and 5.5 billion years, respectively, for β Gem, μ Leo, and α Tau. Thus, the first two giants with their relatively large masses of 2.3 and 1.6 $M_\odot$, turned out to be significantly younger than the sun, while the age of the colder and less massive ($M = 1.2 M_\odot$) giant α Tau is probably comparable to that of the sun.

As for the K5-giant α Tau (Aldebaran), it should be noted that in the BSD it was included in the group of M-giants. Given the significantly lower (compared to our other two giants) acceleration of gravity $\log g = 1.2$ (Table 3) and the comparatively large radius $R \approx 46 R_\odot$, we may assume that this star is of an intermediate type between the K5-giants and the K5-supergiants.

An interesting question: how far from Aldebaran are its planets? That distance is about 7 stellar radii. This may be large enough to prevent the central star's (a supergiant?) having an influence on a planet. This interesting question requires further study.

## 4. Analysis of the Fe I lines: technique

**4.1. Observational data. Measurement of equivalent widths.** For analyzing the Fe I lines we used the same observational spectra for β Gem, μ Leo, and α Tau as in the BSD. The observations for the "benchmark stars" are described in Ref. 26 and the spectra themselves are given in ftp://cdsarc.u-strasbg.fr/pub/cats/J/A%2BA/566/A98/fits/norm/. The observations were carried out on different telescopes with different spectrographs. Our three giants had the following characteristics [26] (resolution and signal/noise ratio *S/N*): for β Gem R=115000 and *S/N* = 287-416 (depending on wavelength), for μ Leo *R* = 81000 and *S/N* = 307-465, and for α Tau, for which two telescopes were used, *R* = 115000 and *S/N* = 47-86 and *R* = 81000 and *S/N* = 209-382. All the spectra obtained in Ref. 26 for the "benchmark stars" were reduced to a single resolution of *R* = 70000. We used these spectra to measure the equivalent widths *W* of the Fe I lines for the three test stars.

In recent years, automatic methods for measuring the equivalent widths of spectrum lines have become popular. Several computer programs for realization of these methods have been published. Here we have compared the results of three programs: ARES [27], TAME [28], and WEISS, the last of which was developed at CrAO by one

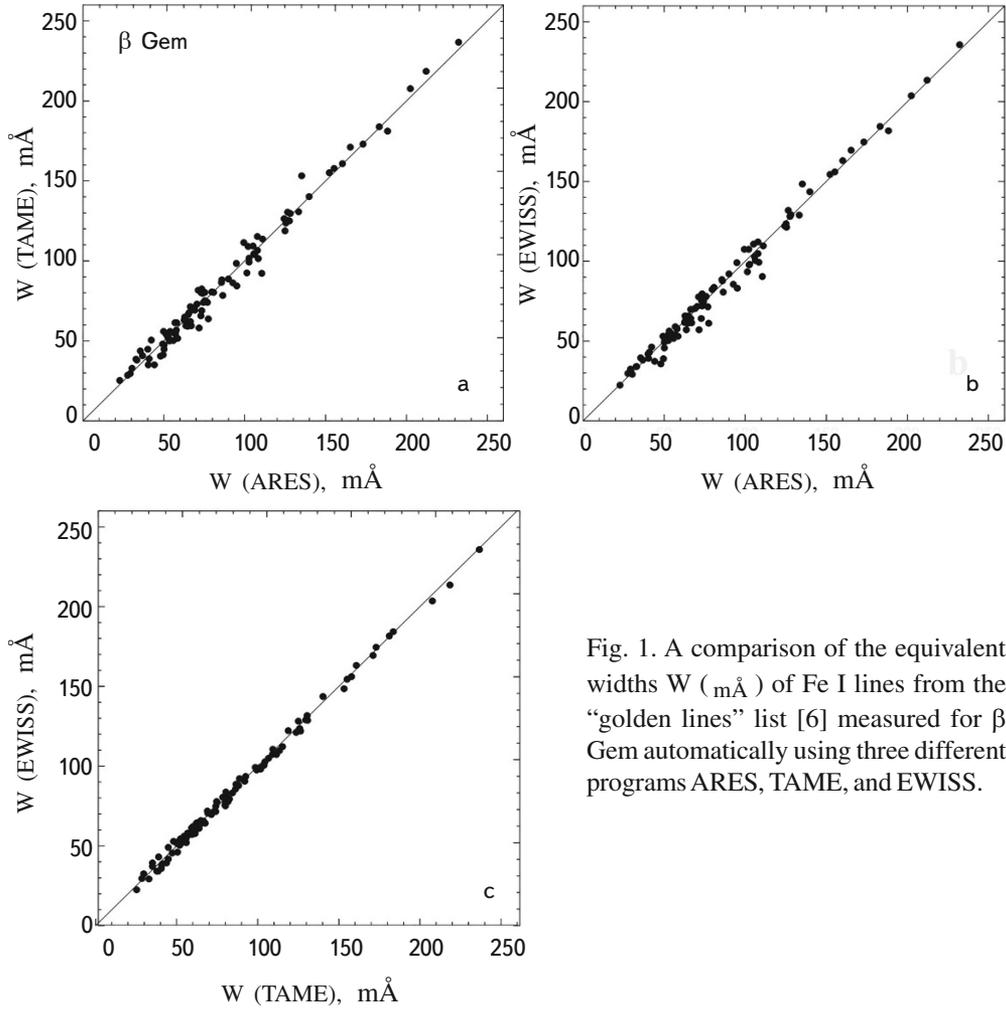

Fig. 1. A comparison of the equivalent widths W ( $_{m\text{Å}}$ ) of Fe I lines from the "golden lines" list [6] measured for β Gem automatically using three different programs ARES, TAME, and EWISS.

of the authors of this paper (D. V. Petrov).

Figure 1 compares the equivalent widths $W$ of Fe I lines in the spectrum of β Gem obtained by these three programs. The "golden lines" selected in the BSD [6] as the most reliable Fe I lines were used. We note that for the FGK-giants, including β Gem and μ Leo, the list of "golden lines" contains 101 Fe I lines, while for the M-giants, to which cold K5-giant α Tau was assigned in Ref. 6, the list contains 21 Fe I lines. The range of wavelengths for the "golden lines" is about 4800-6800 Å and in the case of α Tau, is 4800-6340 Å.

Figure 1 shows that there are no systematic differences between the measurements based on ARES, TAME, and EWISS, but the random dispersion differs significantly: it is higher in the TAME-ARES (Fig. 1a) EWISS-ARES (Fig. 1b) comparisons and considerably smaller in the EWISS-TAME comparison (Fig. 1c). In fact, the mean square deviation is 5.7 and 5.2 mÅ in the first two cases and only 2.7 mÅ in the third. We note that the correlation coefficients are 0.992, 0.993, and 0.998, respectively, in Figs. 1a, 1b, and 1c.

The TAME program was published 5 years after the ARES program. On comparing the methods of these two programs, which are discussed in some detail in Refs. 27 and 28, we concluded that the TAME program is more precise than ARES, since it uses better algorithms for processing the observed spectra. As for the EWISS program, the algorithms used in it differ from both ARES and TAME; nevertheless, we sure that they are quite accurate, as confirmed by the good agreement with the results of the TAME program (Fig. 1c).

Later, for determining the metallicity coefficient [Fe/H] and the microturbulence parameter $V_t$, we relied on the equivalent widths of the Fe I lines obtained using our program EWISS.

**4.2. Non-LTE corrections to the iron abundance.** We determined the metallicity index [Fe/H] along with the microturbulence parameter $V_t$ by analyzing Fe I lines using the traditional assumption of LTE (local thermodynamic equilibrium). However, when analyzing [Fe/H] and $V_t$ one should first estimate how deviations from LTE may affect the results. This later became important when an ambiguity was discovered in the results (see below).

We determined the non-LTE corrections $\Delta_{NLTE}$ for the iron abundance using the calculations of Mashonkina, et al. [29], for cold giants and supergiants with parameters $T_{eff}$ = 4000–5000 K, $\log g$ = 0.5–2.5, and [Fe/H] = –4.0–0.0. That paper makes it possible to obtain, from specified parameters $T_{eff}$, $\log g$, and [Fe/H], the corrections $\Delta_{NLTE}$ for a number of lines that we used from the "golden lines" list. There were almost 30 of these lines. We note that for the giant β Gem, with $\log g$ = 2.85, which lies somewhat outside the range of $\log g$ indicated above, we took $\Delta_{NLTE}$ for estimating $\log g$ = 2.5. For a similar reason, in the case of the giant α Tau, instead of the found value of $T_{eff}$ = 3920 K, for the estimate of $\Delta_{NLTE}$ it was assumed that $T_{eff}$ = 4000 K. Since the corrections $\Delta_{NLTE}$ were small (see below),

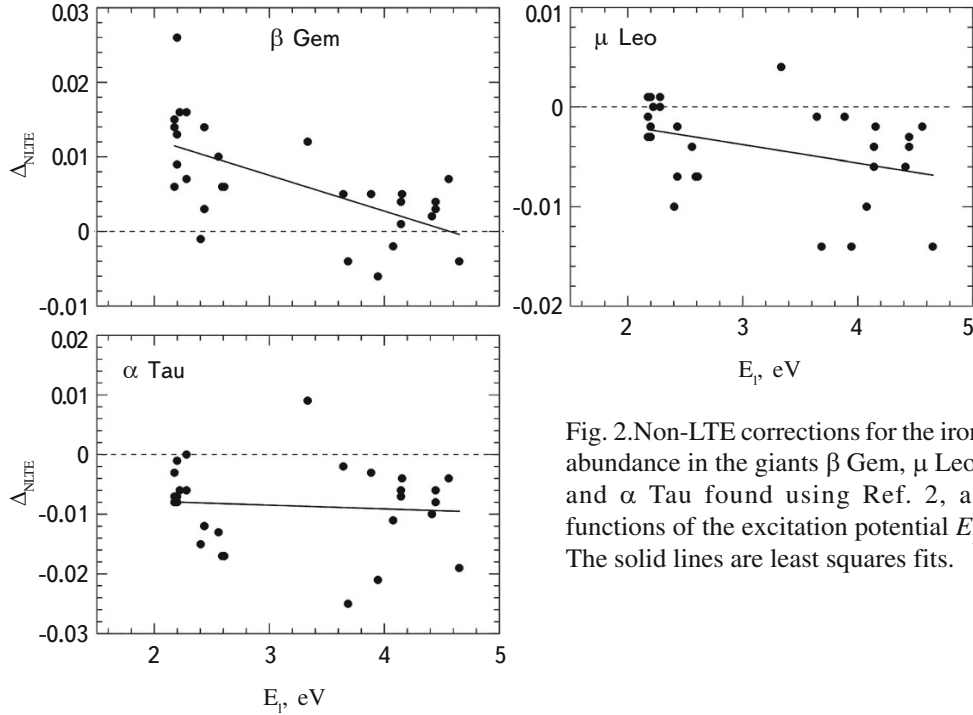

Fig. 2. Non-LTE corrections for the iron abundance in the giants β Gem, μ Leo, and α Tau found using Ref. 2, as functions of the excitation potential $E_1$. The solid lines are least squares fits.

this substitution is of no fundamental importance.

In Fig. 2 the corrections $\Delta_{NLTE}$ are plotted as functions of the excitation potential for the lower level $E_l$. As will be noted below, the parameter $E_l$ plays an important role in analyzing the Fe I lines. A clear trend in $\Delta_{NLTE}$ with increasing $E_l$ can also be seen in the case of the hotter giant β Gem. This trend is less distinct in the case of μ Leo and essentially vanishes for the coolest giant α Tau.

Generally, the corrections $\Delta_{NLTE}$ turned out to be very small. We have evaluated their effect on the determination of [*Fe/H*] and $V_t$ for the example of the giant β Gem, where the dependence of $\Delta_{NLTE}$ on $E_l$ is most distinct. It appears that the non-LTE effects in the Fe I lines lead to change in [*Fe/H*] by only 0.01 dex, and in $V_t$ by 0.04 km/s. Thus, the non-LTE corrections can be neglected for the Fe I lines. We note that this conclusion holds for K-giants with normal or slightly reduced metallicity. For giants with low metallicity [*Fe/H*]~–2 and –3, however, the non-LTE corrections become more significant [29].

## 5. Analysis of Fe I lines: determination of the parameters [*Fe/H*] and $V_t$.

As noted above, we relied on the list of "golden lines" from the BSD for determining the amount of iron logε(Fe) (or the value of [*Fe/H*]) and the microturbulence parameter $V_t$ based on the Fe I lines. Our analysis showed that, besides the dependence of [*Fe/H*] and $V_t$ on the equivalent width $W$, there is a less obvious dependence on the excitation potential $E_l$ for the lower level.

**5.1. The role of the excitation potential $E_l$ of the lower level.** We determined the iron abundance logε(Fe) and the microturbulence parameter $V_t$ simultaneously from the Fe I lines using the traditional method which essentially assumes that there must be no dependence of logε(Fe) on the equivalent width $W$ for the found values of logε(Fe) and $V_t$. In other words, for a given value of $V_t$ the weak and relatively strong Fe I lines should yield the same abundance logε(Fe) on the average.

We began by asking: how do the values of logε(Fe) found from individual Fe I lines depend on the assumed microturbulence parameter $V_t$? As an answer to this question we propose examining Fig. 3, which shows the difference in the iron abundance Δlogε(Fe) for the giant β Gem in the cases $V_t$ = 1 and 2 km/s as functions of the equivalent width $W$. Note that we have not chosen these two values of $V_t$ randomly; according to the BSD, $V_t$ = 1-2 km/s is typical for most K-giants (see Fig. 3 of Ref. 6).

In order to cover a sufficiently wide range of equivalent widths W, in constructing Fig. 3 we chose lines of Fe I in the wavelength range of interest to us from the VALD3 data base [30] that overlapped as little as possible. This choice involved calculating synthetic spectra in two variants for each Fe I line: (1) only for the given choice of Fe I line neglecting nearby lines of other elements and (2) including other lines. If the difference between the two variants in the calculated values of $W$ exceeded 5%, these lines were rejected. After this kind of choice, 873 Fe I lines remained in the list; they are all shown in Fig. 3.

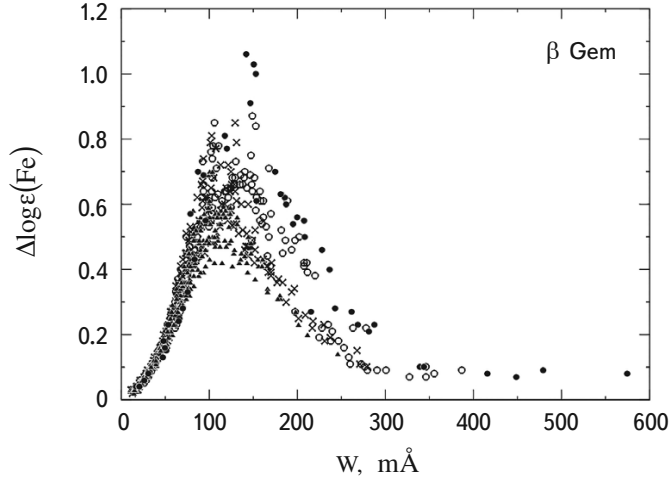

Fig. 3. The difference in the iron abundance $\log\varepsilon(Fe)$ for the giant β Gem between the cases $V_t=1$ and 2 km/s as functions of equivalent width $W$. The lines with different excitation potentials $E_l$ are indicated by the following symbols: $E_l=0.0-1.5$ eV solid circles; $E_l=1.5-3.0$ eV hollow circles; $E_l=3-4$ eV crosses; $E_l=4-5$ eV solid triangles.

The synthetic spectra were calculated using the program SME [31]. We found the iron abundance $\log\varepsilon(Fe)$ from the measured equivalent widths $W$ using the program WIDTH9 [32], as modified by V. V. Tsymbal. Recall that the metallicity index $[Fe/H]$ is related to $\log\varepsilon(Fe)$ by the formula $[Fe/H]=\log\varepsilon(Fe)-\log\varepsilon_\odot(Fe)$, where $\log\varepsilon_\odot(Fe)=7.50$ for the sun [5].

First of all, Fig. 3 attracts attention by the strong dependence of the difference $\Delta\log\varepsilon(Fe)$ on the equivalent width $W$. While for weak lines with $W\sim 10$ mÅ, $\Delta\log\varepsilon(Fe)$ is close to zero (for these lines it is known that the elemental abundance is essentially independent of the choice of $V_t$), it increases rapidly as $W$ rises to 100 mÅ. Near $W=150$ mÅ a maximum of ~1.0 dex is reached and with further increases in $W$, $\Delta\log\varepsilon(Fe)$ shows a decrease, so that lines with $W>300$ mÅ become generally insensitive to changes in $V_t$.

Over the range of $W$ from 100 to 300 mÅ there is a large dispersion in the points on the ordinate, especially in the region of the maximum near $W=150$ mÅ. While replacing $V_t=1$ km/s by $V_t=2$ km/s at the lower points here yields an increase in the abundance of iron $\Delta\log\varepsilon(Fe)$ by ~0.4 dex, at the highest points $\Delta\log\varepsilon(Fe)\sim 1.1$. Thus, the difference in $\Delta\log\varepsilon(Fe)$ between the upper and lower points near $W=150$ mÅ reaches 0.7 dex. What is the cause of such a wide dispersion?

We found that the main reason for the dispersion is the difference in the excitation potentials $E_l$ of the lower level of the Fe I lines examined here. In order to illustrate this conclusion we separated all the lines into four groups according to the potentials $E_l$. They are indicated by different symbols in Fig. 3. We can see that the resulting four sequences of points separate quite distinctly over the interval of $W$ from 120 to 300 mÅ. In particular, the separation between Fe I lines with the lowest potential $E_l = 0.0 - 1.5$ eV (solid circles) and those with the highest potential $E_l = 4 - 5$ eV (solid triangles) is clearly visible. The largest upward jump for the points at $W = 150$ mÅ corresponds to Fe I lines with very low potentials $E_l \approx 0.1$ eV. It is interesting that a separation as obvious as this is no longer observed for $W \leq 150$ mÅ; here the different symbols seem to be mixed.

We have plotted similar dependences of $\Delta\log\varepsilon(\text{Fe})$ on $W$ for the colder K-giants μ Leo and α Tau. These turned out to be very similar to those in Fig. 3 for β Gem. In particular, there is a maximum near $W \approx 100 - 200$ mÅ, and the difference $\Delta\log\varepsilon(\text{Fe})$ between lines with different potentials $E_l$ near $W \sim 150$ mÅ is also about 0.7 dex.

We have concluded that when determining the parameters $\log\varepsilon(\text{Fe})$ and $V_t$ for K-giants, one should probably not only account for the difference in the observed equivalent widths $W$ of the Fe I lines, but also for the difference in the excitation potential $E_l$ of the lower levels of these lines, especially for lines with widths $W$ between 100 and 300 mÅ. Thus, the problem of the excitation potential $E_l$, which showed up in analyzing Fig. 3, may be of highly practical significance.

**5.2. The nonuniform distribution of Fe I lines with respect to the potential $E_l$ in the "golden lines" list: the example of β Gem.** As noted above, for determining [Fe/H] and $V_t$ we used the "golden lines" selected in the BSD [6] as the most reliable Fe I lines. They were used in Ref. 6 for estimating [Fe/H] and $V_t$, but there the possible effect of differences in the excitation potential $E_l$ between the lines was neglected. Can this effect be ignored? We have looked at this question for the example of the giant β Gem.

We analyzed the distribution of the excitation potential $E_l$ for all the "golden lines" (101 lines) of β Gem as a function of $W$. This distribution is shown in Fig.4, which shows, in particular, that the equivalent widths $W$ of these lines vary over a range from 30 to 240 mÅ.

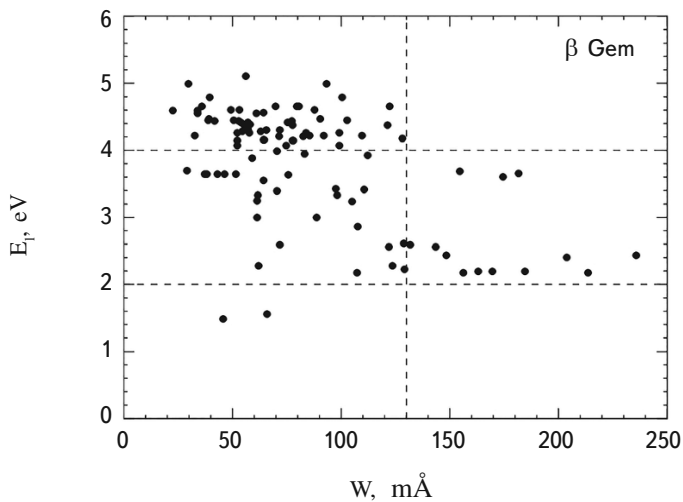

Fig. 4. Fe I lines from the "golden lines" list for the giant β Gem: excitation potential $E_l$ as a function of equivalent width $W$.

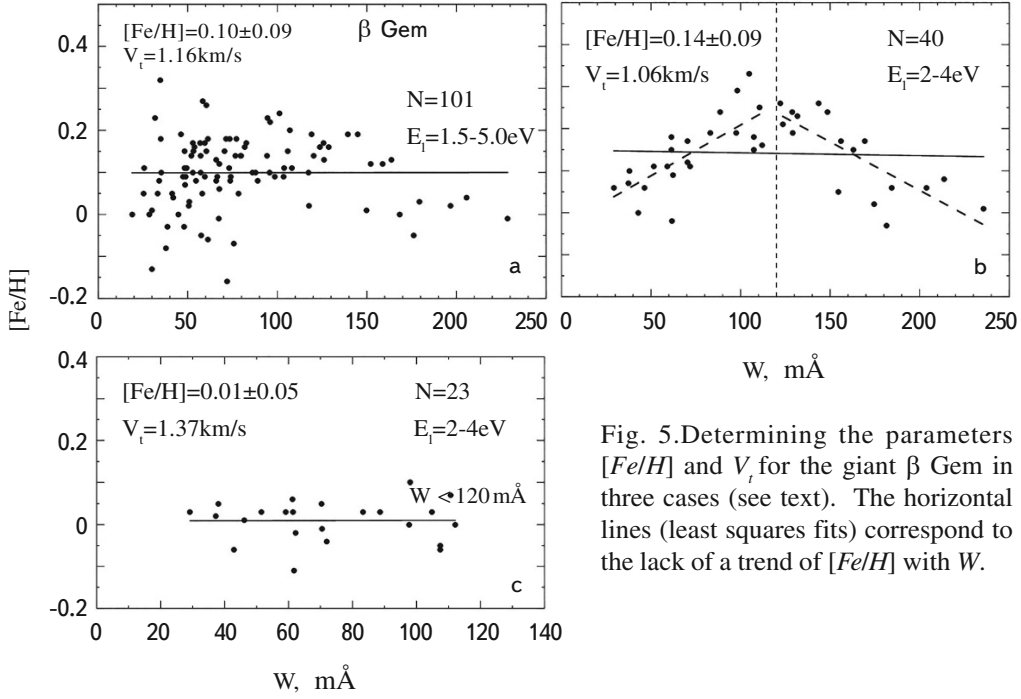

Fig. 5. Determining the parameters [*Fe/H*] and $V_t$ for the giant β Gem in three cases (see text). The horizontal lines (least squares fits) correspond to the lack of a trend of [*Fe/H*] with *W*.

  Figure 4 implies that the distributions were substantially different for relatively strong and weak lines. In fact, for lines with $W = 130-240\,\text{mÅ}$ (the region sensitive to $E_l$) all the values of $E_l$ lie within a fairly narrow interval between 2 and 4 eV (indicated by the two horizontal dashed lines in Fig. 4). On the other hand, for weaker lines with $W \sim 130\,\text{mÅ}$, the values of $E_l$ vary over a wider range from 1.5 to 5 eV; then most of the lines (65%) have high values of 4-5 eV. Thus, the relatively weak and strong lines in the list of "golden lines" have systematic differences in their values of $E_l$. Since a determination of [*Fe/H*] and $V_t$ by the traditional method is based precisely on comparing relatively faint and strong lines, ignoring the systematic differences in $E_l$ between these lines may lead to errors in estimating [*Fe/H*] and $V_t$.

  This is illustrated by Figs. 5, a, b, and c which show [*Fe/H*] and $V_t$ determined for β Gem in three cases corresponding different numbers N of Fe I lines that were used. In all cases the standard method of determining [*Fe/H*] and $V_t$ was used. The main requirement is that, for the chosen value of $V_t$ thereshould not be a global trend in the abundance of iron, i.e., there are values of [*Fe/H*] within the entire range of W from 30 to 240 mÅ.

  Figure 5a shows the results of the analysis for all the "golden rule" lines chosen in the BSD for FGK-giants; here the entire number of lines is $N = 101$. The horizontal straight least squares fit line shows that for $V_t = 1.16$ km/s and [*Fe/H*] $= 0.10\pm0.09$, the above condition is satisfied; i.e., there is no trend in [*Fe/H*] with increasing *W*.

  The rather large dispersion in the points around the horizontal line in this figure is noteworthy. It is especially large for relatively faint lines with $W \leq 100$ mÅ and exceeds ±0.2 dex here.

In order to eliminate the systematic differences between the relatively weak and strong lines, we analyzed only those lines from the "golden lines" list that fall within the band $E_l = 2\text{-}4$ eV (see Fig. 4). The results are shown in Fig. 5b. Of the 101 lines, 40 are left here. A least squares method yields these results (the horizontal line in Fig. 5b): $V_t = 1.06$ km/s and $[Fe/H] = 0.14 \pm 0.09$; they differ little from the previous case (Fig. 5a).

An important fact shows up in Fig. 5b: there is no unique dependence of $[Fe/H]$ on $W$ which might seem suitable for all lines in the range of $W$ from 30 to 240 mÅ. Instead, two different dependences are observed: one for relatively weak and the other for relatively strong lines, with a boundary at approximately $W = 120$ mÅ (the vertical dashed line in Fig. 5b). For clarity these two curves for the points with $W < 120$ mÅ and $W > 120$ mÅ are shown in Fig. 5b by two different dashed curves (least squares fits). It turns out that for lines with $W < 120$ mÅ and with $W > 120$ mÅ the traditional method gives different pairs of numbers $[Fe/H]$ and $V_t$. Such a clear difference between the Fe I lines with $W < 120$ mÅ and $W > 120$ mÅ requires an explanation.

Figure 5c shows only relatively weak lines (a total of 23) with $W < 120$ mÅ. These lines yielded values of $V_t = 1.37$ km/s and $[Fe/H] = 0.01 \pm 0.05$; they are significantly different from those found in the two previous cases. The sharp reduction (by more than a factor of two) in the dispersion in the points compared to the case for lines with $W < 120$ mÅ in Fig. 5a is noteworthy.

In the analysis of Fig. 3 in Section 5.1 it was noted that for lines with $W \leq 100$ mÅ, no clear difference between lines with different potentials $E_l$ is observed. Then the question arises of whether Fig. 5c changes if, instead of the 23 lines in the limited range of $E_l = 2\text{–}4$ eV, all the lines with $W < 120$ mÅ over the entire range of $E_l$ from 1.5 to 5 eV (a total of 81 lines) are shown. We got the following answer: on one hand, the found values of $V_t = 1.31$ km/s and $[Fe/H] = 0.04 \pm 0.09$ were very close to those shown in Fig. 5c. On the other hand, the dispersion in the points near the average of $[Fe/H]$ changed significantly; in fact, while it is within ±0.1 dex in Fig. 5c, in the latter case it increased to ±0.3 dex. Thus, in order to enhance the accuracy of the values of $[Fe/H]$ and $V_t$ for K-giants similar to b Gem, even in the case of relatively weak Fe I lines we recommend using only the lines with potentials $E_l = 2\text{–}4$ eV from the "golden lines" list.

If, finally, we examine only the relatively strong lines with $W > 120$ mÅ, this leads to a further reduction in $V_t$ and $[Fe/H]$ compared to Figs. 5b and 5c; specifically, $V_t = 1.63$ km/s and $[Fe/H] = -0.15 \pm 0.03$. These last values differ by –0.25 dex for $[Fe/H]$ and by 0.47 km/s for $V_t$ from the original values obtained with all the lines from the "golden lines" list (Fig. 5a).

Thus, based exclusively on Fe I lines from the "golden lines" list, we have four variants of the pair $[Fe/H]$ and $V_t$ that differ in the choice of intervals for the parameters $E_l$ and $W$. The results are collected in Table 4. This implies that, first of all, a determination of $[Fe/H]$ and $V_t$ depends significantly on range of $E_l$ being considered. Second, the values of $[Fe/H]$ and $V_t$ are different if we examine the relatively weak and strong lines (i.e., lines with equivalent widths $W < 120$ mÅ and $W > 120$ mÅ) separately.

TABLE 4. Four Variants of a Determination of the Parameters [*Fe/H*] and $V_t$ for the Giant β Gem Based on Fe I lines from the "golden lines" list

| Variant | Number of lines | $E_p$, eV | W, mÅ | [*Fe/H*] | $V_t$, km/s |
|---------|-----------------|-----------|---------|---------------|-------------|
| a | 101 | 1.5 - 5.0 | 20 - 230 | 0.10 ± 0.09 | 1.16 |
| b | 40 | 2 - 4 | 20 - 230 | 0.14 ± 0.09 | 1.06 |
| c | 23 | 2 - 4 | <120 | 0.01 ± 0.05 | 1.37 |
| d | 17 | 2 - 4 | >120 | -0.15 ± 0.03 | 1.63 |

**5.3. Analysis of Fe I lines for μ Leo and α Tau.** We repeated all of the same analysis for the 101 Fe I lines from the "golden lines" list discussed above for β Gem for the case of the colder giant μ Leo. The results of this analysis turned out to be very similar to those for β Gem. In particular, when lines within a limited range of excitation potentials $E_l = 2$–4 eV, as in Fig. 5b, were examined, two different dependences were obtained for the relatively faint and strong lines. Here, however, the separation line lies roughly at $W = 170$ mÅ, rather than $W = 120$ mÅ. This is fully explainable, since, because of the lower temperature $T_{eff}$, the same Fe I lines in the spectrum of µ Leo have higher $W$. Figure 6 (top panel) shows the dependence of [Fe/H] on $W$ for the relatively faint lines with $W < 160$ mÅ (22 lines) in the case of µ Leo. Here values of $V_t = 1.90$ km/s and [Fe/H] = 0.18 were obtained. As for β Gem, they differ significantly from those found for µ Leo compared to the original analysis of 101 lines from the "golden lines" list: $V_t = 1.60$ km/s and [Fe/H] = 0.33.

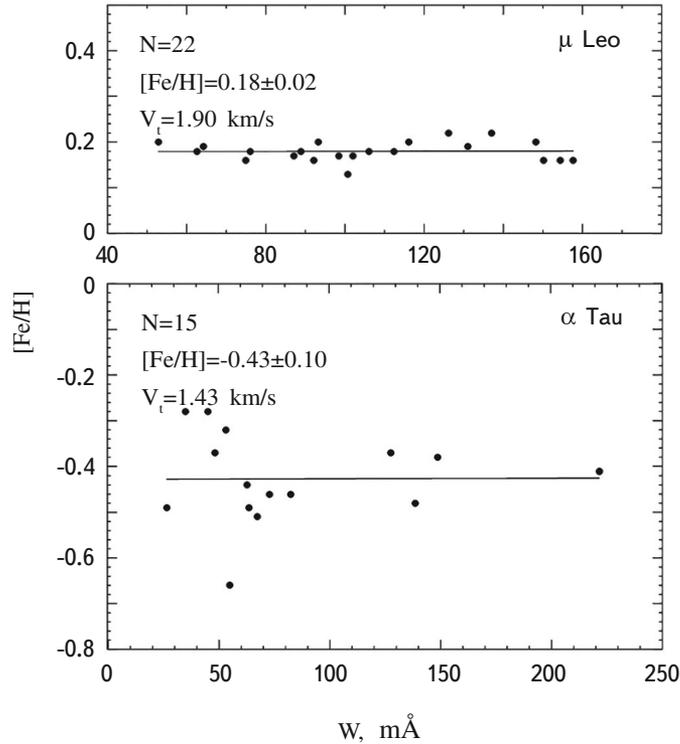

Fig. 6. Determining the parameters [*Fe/H*] and $V_t$ for the giants m Leo and α Tau. The horizontal least squares fits correspond to the absence of a trend of [*Fe/H*] with W.

α Tau, the coolest of the three giants discussed here, is a special case. In the BSD it is included in the group of M-giants and its "golden lines" list in Ref. 6 is substantially shorter compared to β Gem and µ Leo (21 lines are shown instead of 101). In this case, however, a systematic discrepancy shows up in the excitation potentials $E_l$ for the relatively strong lines ($W > 80$ mÅ) and the relatively weak ones ($W < 80$ mÅ). In fact, all 5 of the relatively strong lines with $W$ between 80 and 220 mÅ have potentials $E_l$ between 2 and 4 eV, while the 16 weak lines with $W$ between 10 and 70 mÅ have $E_l$ between 3 and 5 eV.

We have excluded the 6 weak lines with potentials $E_l = 4.4$–5.0 eV from the bottom frame of Fig. 6 for α Tau. The remaining 15 lines with $E_l = 2.0$–4.4 eV yielded $V_t = 1.43$ km/s and [Fe/H] = –0.43 ± 0.10. If we retain the 8 lines within the narrower range of $E_l = 2$–4 eV, we obtain very similar values: $V_t = 1.35$ km/s and [Fe/H] = –0.38 ± 0.12. Note that for all the "golden lines" (21 lines) we originally obtained $V_t = 1.39$ km/s and [Fe/H] = –0.39 ± 0.11 for α Tau. We can see that all three variants actually yield the same (within the measurement error) values of [Fe/H] and $V_t$. Thus, for the coolest K-giant, α Tau, the role of systematic differences in the excitation potential $E_l$ between the Fe I lines was not as significant as for the hotter giants β Gem and µ Leo.

**5.4. Comparison with the BSD results.** In comparing our estimates of [Fe/H] and $V_t$ with the BSD data, we should note, first of all, that different models of atmospheres are used in Refs. 6-8 and in our work: in the first case, the MARCS models and in our calculations, the ATLAS9 models (their current versions are discussed in Ref. 33). As our calculations showed, however, in the case of K-giants the difference in these atmospheric models has little effect on a determination of the amount of iron based on Fe I lines.

A more serious comment concerns the BSD data introduced in Ref. 6. There the parameter [Fe/H] was determined by different groups using 7 different methods (6 methods for α Tau). These methods yielded a noticeable dispersion (see Table 2 in Ref. 6). In particular, for β Gem the variations in [Fe/H] ranged from 0.00 to +0.24 (i.e., the dispersion = 0.24 dex), for µ Leo, from +0.23 to +0.50 (dispersion = 0.27), and for α Tau, from -0.43 to -0.12 (dispersion = 0.31 dex). These data can serve as a good illustration of the accuracy of modern data on the metallicity parameter [Fe/H] for K-giants.

In Table 5 we compare the values of [Fe/H] and $V_t$ obtained using relatively weak Fe I lines with the averaged BSD data [6]. It is clear that the agreement is quite good, despite the above noted ambiguity in the determination of [Fe/H] and $V_t$. Both data sets confirm that the K-giant Gem has a metallicity close to that of the sun; the giant µ Leo has a slightly higher metallicity $[Fe/H] \approx 0.2$ dex, while the coolest giant α Tau definitely manifests a reduced metallicity of $[Fe/H] \approx -0.4$ dex.

TABLE 5. Our Values of [*Fe/H*] and $V_t$ Compared with Data from BSD [6]

| Parameter | Source | β Gem | μ Leo | α Tau |
|---|---|---|---|---|
| [*Fe/H*] | our work | 0.01±0.05 | 0.18±0.08 | -0.43±0.10 |
|  | BSD | 0.13±0.16 | 0.25±0.15 | -0.37±0.17 |
| $V_t$, km/s | our work | 1.4±0.2 | 1.9±0.3 | 1.4±0.3 |
|  | BSD | 1.3±0.2 | 1.3±0.3 | 1.6±0.3 |

## 6. Discussion

With regard to the evolutionary status of the K-giants studied here there is some interest in whether they have passed through Deep Convective Mixing (DCM) or, in the traditional terminology, the "First Dredge Up." The data of Refs. 34-36, shown in Table 6, can serve as an answer. They definitely indicate that all three stars have already passed the DCM phase. The proof is the very low (compared to the sun and young stars) ratio of the carbon isotopes, $^{12}C/^{13}C = 10 - 18$ (for the sun $^{12}C/^{13}C = 89$), as well as a substantially reduced ratio of the oxygen isotopes, $^{16}O/^{17}O = 240 - 1670$ (for the sun $^{16}O/^{17}O = 2632$).

How well do these values fit the predictions of theory? The role of stellar rotation in the evolution of stars is now well known. For example, stellar model calculations with rotation have made it possible to explain

TABLE 6. Ratios of the Isotopes of Carbon and Oxygen in the Atmospheres of the Three K-Giants Studied Here and the Sun

| Star | $^{12}C/^{13}C$ | $^{16}O/^{17}O$ | Source |
|---|---|---|---|
| β Gem | 18 | 240±60 | [34] |
| μ Leo | 18 ± 3 | 325±100 | [35] |
| α Tau | 10 ± 2 | 1670±550 | [36] |
| Sun | 89.4 | 2632 | [5] |

**quantitatively** the long known "nitrogen-oxygen" anticorrelation for AFG-supergiants [37], as well as the observed change in the N/O ratio at the end of the MS stage for B-stars with masses from 5 to 20 $M_\odot$ [38]. These effects depend significantly on the mass M and on the initial rotation velocity $V_0$. Model calculations of rotating stars with masses from 15 to 1.7 $M_\odot$ have shown that, because of rotation, the abundances of the light elements C, N, and O on the surfaces of these stars can change significantly already by the end of the MS stage, and that these changes intensify after completion of the DCM stage [39].

We are interested in giants with low masses $M \approx 1-2\,M_\odot$. Abia, et al. [36] assume that mixing of these stars by rotation is ineffective because of the low rate of mixing. On analyzing the observed isotope ratios of carbon and oxygen for two nearby K-giants, Aldebaran and Arcturus ($M \approx 1-2\,M_\odot$ for both of them), they proposed the idea of additional, nonconvective mixing ("extra-mixing") in order to explain the observed ratios of these isotopes. Their calculations including "extra-mixing" for models with $M = 1.08$, 1.2, and 1.3 $M_\odot$ showed that after DCM the ratio $^{12}C/^{13}C$ may take values from 3 to 17, which is fully consistent with the minimal observed values of $^{12}C/^{13}C$ for the K-giants. As to the calculations for $^{16}O/^{17}O$, values from 1380 to 4830 were obtained; Table 6 shows that these estimates are excessive in the case of the K-giants β Gem and μ Leo. Thus, the problem cannot be said to be fully solved, at least for the case of the ratio $^{16}O/^{17}O$.

The ambiguity we have discovered in determining the parameters $V_t$ and [Fe/H] by analysis of Fe I lines, which remains even after elimination of the systematic differences between relatively weak and strong lines (Fig. 5b) in the excitation potentials $E_l$, requires an explanation. On one hand, the question of how to explain this ambiguity theoretically is interesting. On the other hand, this problem has a purely practical significance because the Fe I lines from the "golden lines" list [6] will undoubtedly be used in other work on the FGK-giants. In this regard, the following question arises: which lines can be recommended for use?

We recommend using the relatively weak Fe I lines ($W < 120\,\text{mÅ}$, $E_l = 2-4$ eV) for the early K-giants with temperatures $T_{eff}$ of roughly 4400-4900 K, such as β Gem and μ Leo. Then for analyzing the abundances of other chemical elements it is also necessary to rely on relatively weak lines with equivalent widths of no more than 100 mÅ when possible. For the coolest K-giants with $T_{eff} = 3900-4000$ K, such as α Tau, where the number of reliable Fe I lines from the "golden lines" list is relatively small and where the difference in the excitation potentials $E_l$ is less significant, it appears that Fe I lines with widths up to W~200 mÅ can be used.

One possible explanation for the ambiguity we have found in determining the parameters $V_t$ and [Fe/H] using the Fe I lines may be the idea of deviations from LTE that have not been taken into account. However, as pointed out above (Section 4.2), the role of deviations from LTE for the Fe I lines was insignificant in the case of the K-giants. Thus, other causes for this ambiguity must be sought.

It might be suggested that the atmospheric models being used are inadequate. As noted above, we relied on the ATLAS9 models and the MARCS models were applied in the BSD papers. For cold K-giants these models may manifest differences in the photometric indices $Q$ and $[c_1]$ (see Ref. 14), but in studies of elemental abundance for these kinds of stars, no significant differences were found between the ATLAS9 and MARCS models. It is important that these are all stationary, one-dimensional atmospheric models. Might it be that they are inadequate for the actual atmospheres of cold K-giants and that here three-dimensional hydrodynamic 3D-models are more appropriate?

As an example we point out a recent study [40] of the K-giant halo HD 122563 with parameters $T_{eff} = 4600$ K, $\log g = 1.6$, and $[Fe/H] \approx -2.5$. It turned out that a 3D-model did not solve all the problems regarding the Fe I lines. A trend in the iron abundance with increasing $W$ leads to the conclusion that even here it is necessary to account for the microturbulence parameter $V_t$, although the first calculations for the sun seemed to show that when a 3D-model is used there is no need to introduce this parameter. In addition, an LTE analysis gives an excessively large difference ~0.4 dex in the abundance of iron between Fe I and Fe II lines. This discrepancy could be attributed to unaccounted-for deviations from LTE for the Fe I lines (they are insignificant for the Fe II lines). According to Ref. 29, however, even for a a metallicity of $[Fe/H] = -3$, the expected non-LTE corrections should not exceed 0.2-0.25 dex (see Fig. 2 in Ref. 29). This example shows that in the case of a complicated atom such as iron, the existing non-LTE calculations are still not entirely adequate, at least for stars with a low metallicity.

It is clear that the transition from one-dimensional models of stellar atmospheres to 3D-models for K-giants requires further study.

## 7. Concluding comments

The adequately high accuracy of our technique for determining the basis parameters of K-giants has been confirmed by illustrating it with the example of three bright, nearby K-giants with planets (β Gem, μ Leo, and α Tau). This confirmation was based on a comparison with high precision data for benchmark stars obtained in the Gaia project.

The effective temperature $T_{eff}$, acceleration $\log g$ of gravity in the atmosphere, mass $M$, and luminosity $L$ found by us were in good agreement with data for the benchmark stars, and the resulting differences are within the limits of error of the measurements.

All three of the giants appear to have passed through the stage of deep convective mixing, as indicated, in particular, by a low carbon isotope ratio $^{12}C/^{13}C$ in their atmospheres.

The theoretical interpretation of the ambiguity in determining the parameters $[Fe/H]$ and $V_t$ for the K-giants remains an open question. In practice, today we could make the following recommendations: when estimating $[Fe/H]$ and $V_t$ for K-giants with effective temperatures $T_{eff}$ from 4400-4900 K (e.g., β Gem and μ Leo), one should rely on Fe I lines with equivalent widths $W < 120$ mÅ and excitation potentials $E_l = 2-4$ eV. When determining the abundances of other chemical elements for these stars, one should also use lines that are as weak as possible with widths $W$ of no more than 100 mÅ. For the coolest K-giants with temperatures $T_{eff} \approx 3900-4000$, such as α Tau (Aldebaran), it is permissible to use Fe I lines with W up to 200 mÅ. In that case, the range of potentials $E_l$ is insignificant.

We thank V. V. Shimanskii for a useful discussion.